\documentclass[aps,pre,longbibliography,notitlepage,floatfix,11pt,tightenlines,nofootinbib,superscriptaddress,showkeys]{revtex4-1}
\usepackage{float}
\usepackage{placeins}  

\usepackage{hyperref} 

\usepackage{bm,amsmath,amssymb,graphicx,stmaryrd,subfigure,xcolor}
\usepackage[abs]{overpic}

\begin{document}

\title{A sequence of elastic patterns in a sheared bent sheet}

\author{D. Gimeno}
    \email{D.Gimeno@sussex.ac.uk}
\author{B. K. Meghwar}
\author{G. Fisher}
\author{R. S. Hutton}
\affiliation{Department of Mechanical Engineering, University of Nevada, 1664  N.\  Virginia  St.\ (0312),  Reno,  NV  89557-0312,  U.S.A.}
\author{E. Hamm}
    \email{luis.hamm@usach.cl}
    \affiliation{Departamento de F\'isica, Facultad de Ciencia, Universidad de Santiago de Chile, Av.\ V\'ictor Jara 3493, Estaci\'{o}n Central, Santiago 9160000, Chile}
\author{J. A. Hanna}
    \email{jhanna@unr.edu}
    \affiliation{Department of Mechanical Engineering, University of Nevada, 1664  N.\  Virginia  St.\ (0312),  Reno,  NV  89557-0312,  U.S.A.}

\begin{abstract} 
	We document a sequence of bifurcations and elastic patterns in sheared bent sheets of intermediate aspect ratio. The sheets undergo inversion of curvature through the passage of localized features, often in S-shaped pairs. Nested force-displacement hysteresis loops provide experimental evidence for snaking. Several mechanisms for coarsening and refinement of the patterns are observed, including splitting, merging, and escape through open boundaries. While most forces, including that required for full snap-through, scale with the length of the sheet, the initial drop in force upon pattern nucleation decreases rapidly with length. 
\end{abstract} 

\date{\today}


\maketitle

\section{Introduction}

Thin elastic plates and shells under loading or confinement are a pattern-forming system with a history of influence on developments in bifurcation theory and nonlinear dynamics, such as the interrelated study of cellular buckling and homoclinic snaking \cite{Hunt99}. 
A few interesting examples of work at the junction of these fields may be found in \cite{Hunt99, Horak06, ThompsonSieber16, KreilosSchneider17, GrohPirrera19, Hunt20, Groh21, GrohHunt21, Ehrhardt20} 
and in other references discussed in the reviews \cite{Thompson15, Champneys19}. Conformational changes in such two-dimensional thin structures are characterized by inhomogeneous and often highly localized elastic deformations \cite{Witten07, BenAmarPomeau97, Chaieb98, CerdaMahadevan98}. Such phenomena may be observed in many settings, including aircraft panels, which often operate in a dynamic post-buckled regime \cite{Spottswood21}. 

In this paper, we report experimental observations of a sequence of bifurcations and patterns previously informally presented in  \cite{crumpledynamics}. The context is one in which a plate or shell structure is inverting the orientation of its curvature; rather than homogeneous snap-through, this process occurs by the passage of localized zones which form the units of the observed patterns \cite{BerkeCarlson68, Kyriakides94, HolstCalladine94, PauchardRica98, MoraBoudaoud02, Das07, RomanPocheau12, Deshpande21, Hutton24}. We here refer to such zones as ``crumples''. 
We explore the sequence of patterns and bifurcations through force-displacement hysteresis loops of various sizes for sheets in a range of intermediate aspect ratios.  These reveal a structure of solution branches consistent with snaking, a phenomenon in which different numbers of unit cells of a pattern can exist in overlapping ranges of parameters (see \cite{Knobloch15, Dawes10} for reviews). We also briefly examine the forces and drops in force associated with these processes. 

This study grew out of an exploration of the behavior of narrow elastic strips \cite{YuHanna19}.  In a previous paper \cite{Hutton24}, we documented when and how snap-through occurs through inhomogeneous deformation of narrow to intermediate aspect ratio plates of different thicknesses, with an emphasis on displacement measurements and simple geometric arguments. 
We observed the nucleation and propagation of both unstable and stable crumples, which mediated the snap-through process.  
A variety of transient events were seen with a fast camera. Stable intermediate buckled states appeared as the aspect ratio increased, and featured two types of crumple pairs, the O-valley and S-ridge. These pairs, to be re-introduced later in the present paper, will appear prominently as we focus on quasistatic progressions through stable states with multiple crumples, and associate these with force measurements.  
Our initial view was that a crumple serves as an increment of curvature reversal, progressing through the sheet to effect a conformational change.  The mobility of, and possible interactions between, crumples provide mechanisms for pattern formation and evolution. 
At some point, it became apparent that the dynamic and often irregular arrays of features we observed could be connected to the static regular patterns of crumple pairs that appear in the buckling of cylinders under quite different boundary conditions \cite{Harris61, Yamaki,  EsslingerGeierBook}. With this hindsight, we attempt to place our findings in the context of the same snaking framework used to understand these earlier studies \cite{KreilosSchneider17, GrohPirrera19, Hunt20, Groh21, GrohHunt21}. Recent developments in soft matter physics have also stressed the importance of O-valleys (these pairs of what we call crumples are themselves called ``crumples'' in \cite{Timounay20}).  Our focus on dynamics in an open-boundary system will lead us to emphasize S-ridges.

While the patterns we discuss resemble those well known in plate and shell buckling settings such as the pressurization and/or axial compression and/or torque of cylinders or spheres, not only the shear-like and partially open boundary conditions we employ, but the \emph{evolution} of the patterns, are of a different type than those previously studied. Rather than sequential nucleation of cells on a static grid 
\cite{Harris61, Almroth64, Tennyson69, EsslingerGeierBook, Marthelot17, KreilosSchneider17, GrohPirrera19, Hunt20, Groh21, GrohHunt21, Timounay20, Rudd23}, 
we observe refining and coarsening (up to complete elimination) of the pattern through events such as splitting or merging, entrance or escape, enabled by mobility of its fully elastic constituent crumples. Some refinement and coarsening were also documented in the early systematic experiments of Yamaki \cite{Yamaki}, but without details of the mechanism; entrance or escape were precluded by closed boundary conditions. An example of some type of merging can be found in Ravulapalli and co-workers \cite{Ravulapalli24}. Evensen \cite{Evensen64} notes that ``the buckling pattern shifted and moved as it developed''.  
The observations reported here of pattern mobility and transitions are both qualitatively different and more extensive. 

This paper is organized as follows. 
Methods and errors are presented in Section \ref{methods}, where we also re-introduce the concept of O-valley and S-ridge pairs.  A qualitative overview of the relevant phenomena, including mechanisms for changes in patterns, is presented in Section \ref{observations} using information from videos. Quantitative and qualitative results from force-displacement loops are presented in Section \ref{results} and Appendix \ref{additionaldata}. 
Discussion follows in Section \ref{discussion}. 

\section{Methods}\label{methods}

\begin{figure}[H]
    \centering
    \includegraphics[width=\textwidth]{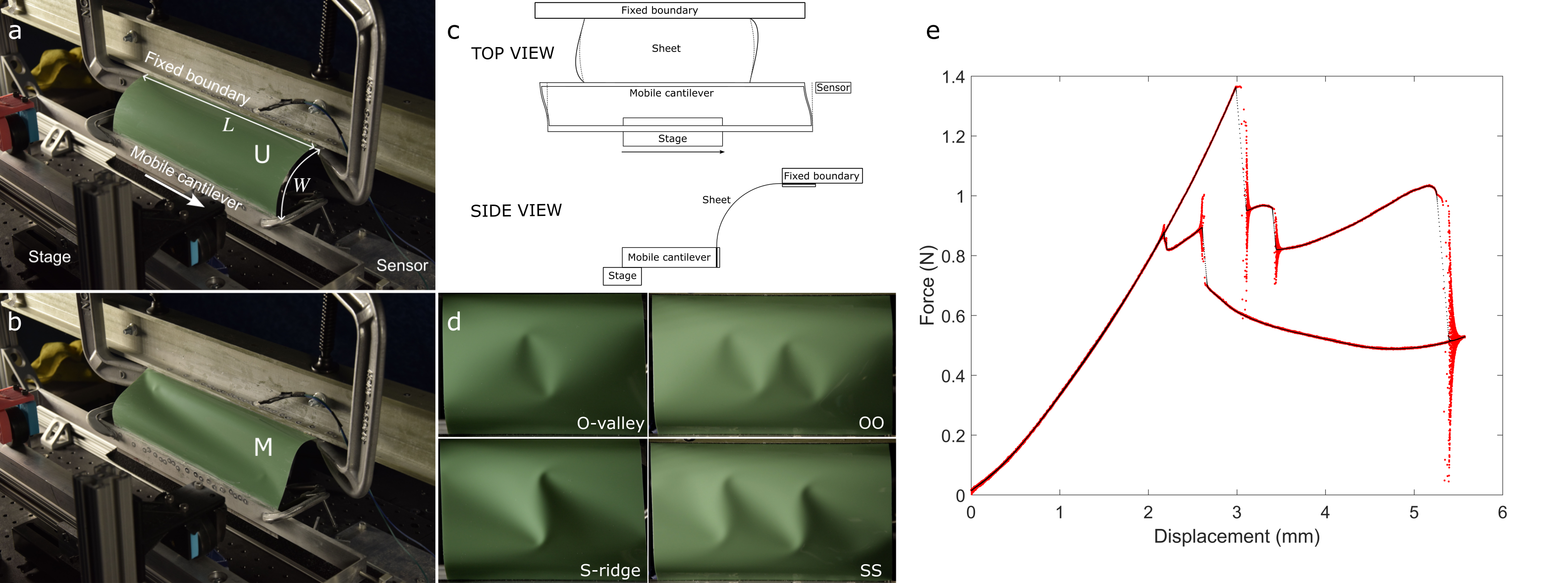}
    \caption{a and b: Experimental setup showing a laterally translating, instrumented boundary, and a 12 inch long ($L$), 4 inch wide ($W$), 0.003 inch thick sheet of polyester (304.8 x 101.6 x 0.0762 mm).  a: The sheet is bent into a quarter-cylinder in the unsheared unimodal U state. b: The lower boundary has moved rightward parallel to the fixed boundary, shearing the sheet into the multimodal M state, in which two small puckers are also visible.  c: Schematic of the setup, top and side views. As the stage moves to the right, the cantilever bends to the left and the sheet deforms.  Dotted lines in the top view indicate the undeformed positions of sheet and cantilever. d: Single O-valley and S-ridge pairs, and examples of patterns showing two pairs of each, OO and SS, in lower aspect ratio sheets. e: Unfiltered (red) and filtered (black) force-displacement loops, traversed clockwise.  The black curve is the same as that in Figure \ref{force_displacement_6in_alt} in Section \ref{results} below.}
    \label{setup}
\end{figure}

Measurements were taken of thin polyester sheets (0.003 $\pm$0.0002 inch / 0.0762 $\pm$0.00508 mm thick shim stock, Artus Corp., Englewood, NJ, USA), with lengths in the range of 5-12 inch (127-304.8 mm), mounted in a quarter-cylinder configuration with 4 inch (101.6 mm) free span (2.546 inch / 64.77 mm radius), their flat sides attached to parallel beams and bent sides free. One beam was fixed and rigid, and the other was a flexible cantilever attached to a mobile translation stage (Thorlabs, Inc., Newton, NJ, USA).  Some observations were also made of longer sheets. As the material is mildly anisotropic, sheets were oriented with their machine direction parallel to the beams. 
Sheets were attached to the beams using rubber adhesive polypropylene cloth mounting tape, then secured with an additional clamping plate and corner clamps to prevent edge peeling. Alignment was achieved with additional positioning stages, and movement of the cantilever boundary was guided by four stabilizers to prevent twisting or deflection that could occur due to the inclusion of a soft pad between the translation stage and cantilever to damp external vibrations.  Still, errors on the order of 1-2 mm (half a percent of the sheet dimensions) are expected, larger than an estimated error of 0.02 inch (0.508 mm) in free span width and length. The samples were quasistatically sheared by laterally displacing the cantilever boundary, which both moves and deflects parallel to the other boundary. The sheets transition, through a sequence of bifurcations and stable patterns, from an initial unimodal configuration (hereafter the ``U'' state), to a final multimodal configuration (``M'' state). The experimental arrangement is photographed and shown schematically in Figure \ref{setup}, with U and M states for a 12 inch (304.8 mm) long sheet.

Two recurring building blocks for patterns are pairs of crumples bound in an ``O-valley'' or ``S-ridge'' arrangement, as named in \cite{Hutton24}. 
 The O-valley pair is akin to the indentation obtained by poking a soda can, with two crumples facing each other along a valley. The S-ridge pair has two crumples joined along a ridge. It should not be confused with the s-shaped curved and highly extended O-valleys seen during the twisting of cylinders in \emph{e.g.} \cite{Harris61, Hamm04}. 
Figure \ref{setup}D shows these pairs as well as examples of patterns with two pairs each. 
When multiple pairs are present, it is often possible to classify the arrays by whether valley or ridge ``bonds'' are more prominent, as reflected in the way pairs are seen to move (for example, there are O-like valleys between S pairs, but the S pairs will move rigidly and independently so that these valleys change length), but there can be subtle transitions between the two, during which there is some potential for ambiguity. These distinctions should become clearer upon viewing the videos introduced in Section \ref{observations}.
Note that this classification as O or S is not based on whether indentations are connected with the open boundaries, and that additional  unpaired crumples are often present on either or both sides of an array.

The cantilever was machined from a 24 inch long, 1 inch thick (609.6 x 25.4 mm) aluminum plate, leaving two thin beams approximately 5.31 inch long and 0.05 inch thick (135 x 1.27 mm).  
A mild iron target was attached to one beam to enable measurement of deflection using an inductive sensor (Baumer Ltd., Southington, CT, USA), attached to a soldered external piece bolted to the cantilever. The stage-cantilever-sensor arrangement was calibrated by pushing against a rigid object, and by pushing against a load cell (Sentran LLC, Ontario, CA, USA). This allowed us to relate the position of the stage, the deflection of the cantilever, and the force applied by the cantilever.  Displacements reported are those of the sheet edge mounted on the deflected end of the cantilever, not the base of the cantilever mounted on the stage. The cantilever is operated in a linear force-deflection regime, with sub-millimeter deflections. From the deadload measurements, we also determined that the stage backlash can be empirically corrected throughout the relevant load regime by applying a 5 micron shift between forward and backward load curves. This is consistent with manufacturer-supplied accuracy and backlash values. Force data from the sensor had a noisy width estimated as 0.008 Newton, and was smoothed using a Gaussian filter (Matlab smoothdata function which has an SD of 1/5 of window, window 50, acquired at 100 Hz) for the purpose of clearer visualization. 
 An example of the qualitative effects of this smoothing is shown in Figure \ref{setup}e.  The effects can also be inferred from the slope artifacts seen on the force-displacement curves, although we note that slightly sloped jumps occur in a system loaded by a compliant structure such as a cantilever (not strictly displacement controlled).  A few small and short-lived events occurring in longer sheets are masked by the rates and filtering used here.  
 Sensor drift of the same order as the noise was occasionally observed over periods spanning tens of minutes. We estimate the associated drift width as 0.008 Newton. The noise and drift error estimates are consistent with conservative bounds on root mean square deviation of unfiltered and filtered signals over tens to hundreds of minutes over the range of voltages used. As these errors are already small with respect to our measurements, we did not attempt to determine lower estimates for \emph{e.g.} low voltages which display lower drift. 
 
%

The stage is moved at a speed of 0.01 mm/s, with accelerations of 1 mm/s$^2$ at the beginning and end of the loops.  Executing a single full loop takes about 12-18 minutes. We zero the sensor (corresponding to the calibration zero) on the cantilever without any sample attached, then attach and clamp a sheet and move the stage until the sensor is zeroed again.  This serves as a reference symmetric unsheared ``U'' state of the sheet, corresponding to the zero of our displacement.  As can be gleaned from the figures, this zeroing was only performed approximately; discrepancies may be as much as 120 microns.  The stage is moved until the sheet reaches some M configuration. The displacement to effect this and other transitions becomes smaller with increasing aspect ratio of the sheet, as it is bounded by the straightening of the long diagonal \cite{YuHanna19}. The stage is programmed to go back and forth between reference U and M positions and various points of interest in loops of different lengths, to explore different branches of equilibria.  Data is acquired after ``warming up'' the system by cycling it around the largest loop four or five times at a speed of the same order of magnitude as that used for data collection. 

Crumples are filmed with a Nikon D5300 camera, with the sheets illuminated from the side.

\subsection{Annotation of the sequence in Video \ref{vidgreen}}
 
In Section \ref{observations}, a video provides a representative qualitative picture of the full forward and backward sequence (for this video, to better capture certain fast motions with a standard camera, the stage was moved at a speed of 0.003 mm/s instead of the 0.01 mm/s used for taking force data).  Instead of directly annotating the video, we annotate a spacetime plot of crumple positions (in pixels), generated as follows.  
We used a code modified from that generated by Gemini \cite{gemini}. 
Crumple positions were detected using several Matlab \cite{matlab} functions.  The code tracked these and produced a video with overlays indicating the crumple outlines and position proxies, which we used to visually validate the process by iteratively changing parameters in the code, as described below. 

Images were converted to grayscale, separated into upper and lower halves by a diagonal line from upper left to lower right cutting through the crumple pairs. Half of the image was grayscale-inverted before both halves were binarized using dynamic thresholds (imbinarize with our inputs) and a region of interest selected. 
Blobs were detected as 4-connected regions (bwlabel) and infilled to be simply connected (imfill), with detection parameters set according to size (bweareopen, usually below 2000px), 
aspect ratio (bounding box horizontal to vertical aspect ratio, below 3), and duration (five frames) thresholds. Noisy artifacts within 10px of the boundaries of the region of interest were also discarded. 
Geometric information was collected (regionprops) from these detected blobs.  
Six additional spurious regions were identified by comparison with the final overlay video and manually discarded; these are short-lived artifacts of lighting appearing during two transitional times in the pattern formation process. 
The top/bottom points of the boundaries (bwboundaries) for upper/lower blobs serve as imperfect proxies for crumple core positions. We report these, unfiltered, in terms of the pixels of the video frames. 

According to Gemini \cite{gemini}, the generation of the overlay video for visual validation uses tracking algorithms constituting a ``greedy nearest neighbor state machine''.  This portion of the code was created in response to prompts such as ``keep track of blobs'', ``I want to see visualization''.  It contains adjustable parameters that affect identification of blobs that appear, disappear, or merge. 
None of this has any effect on the spacetime plot.


We emphasize that this process was used only to aid in annotation of a sequence of events in a video by creating a spacetime plot. We used the entire video to validate the qualitative accuracy of the positions. No quantitative analysis is intended.

\section{Preliminary observations}\label{observations} 

Let us first introduce the phenomena of interest through observations. 
 An organized sequence of dynamics can be seen in Videos \ref{vidblue}-\ref{vidgreen}, in which long (high aspect ratio) bent sheets are subject to lateral boundary displacement, resulting in a sequence of quasi-1D patterns of crumples.  Note that the deformation shown in V1 is due to boundary displacement in the opposite direction as in V2 and all the experiments in this paper; it is effectively a mirror image of the behavior. 
 The sheet is also about twice as large and thick, with longer aspect ratio, and the boundary is moved more than two orders of magnitude faster, than in the experiments.  Video \ref{vidgreen} involves the same conditions as our highest aspect ratio experiments, detailed above in Section \ref{methods}. 
 
\begin{video}[H]
	\centering
	\includegraphics[width=0.25\textwidth]{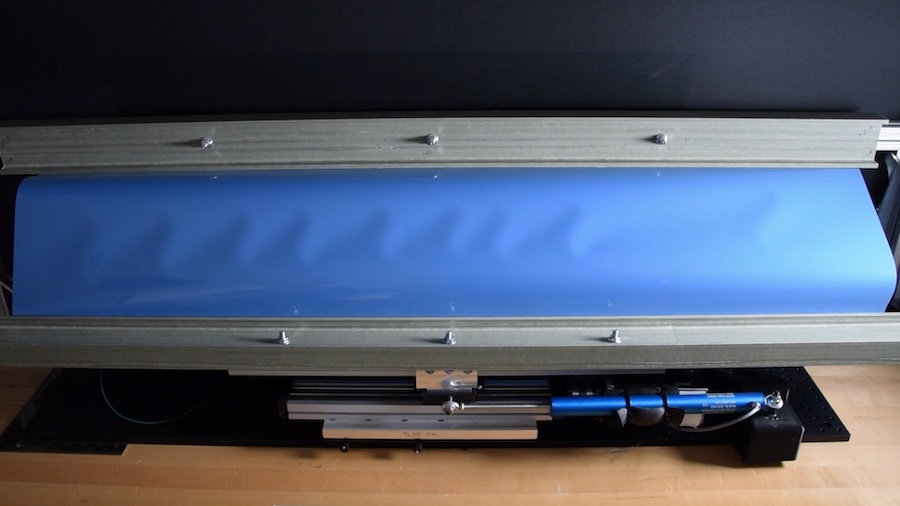}
	\setfloatlink{http://doi.org/10.5281/zenodo.18487904} 
	\caption{
Patterns in 0.005 inch (0.127 mm)  polyester shim stock, 48 inches (1219.2 mm) long, bent into a quarter-cylinder with 10 inches (254 mm)  free span, clamped with lower boundary moving laterally to the left and then to the right. The boundary is moved more than two orders of magnitude faster than 
in experiments used for data.  Excerpt from \cite{crumpledynamics}.
}\label{vidblue}
\end{video}

\begin{figure}[h]
    \centering
    \includegraphics[width=0.9\textwidth]{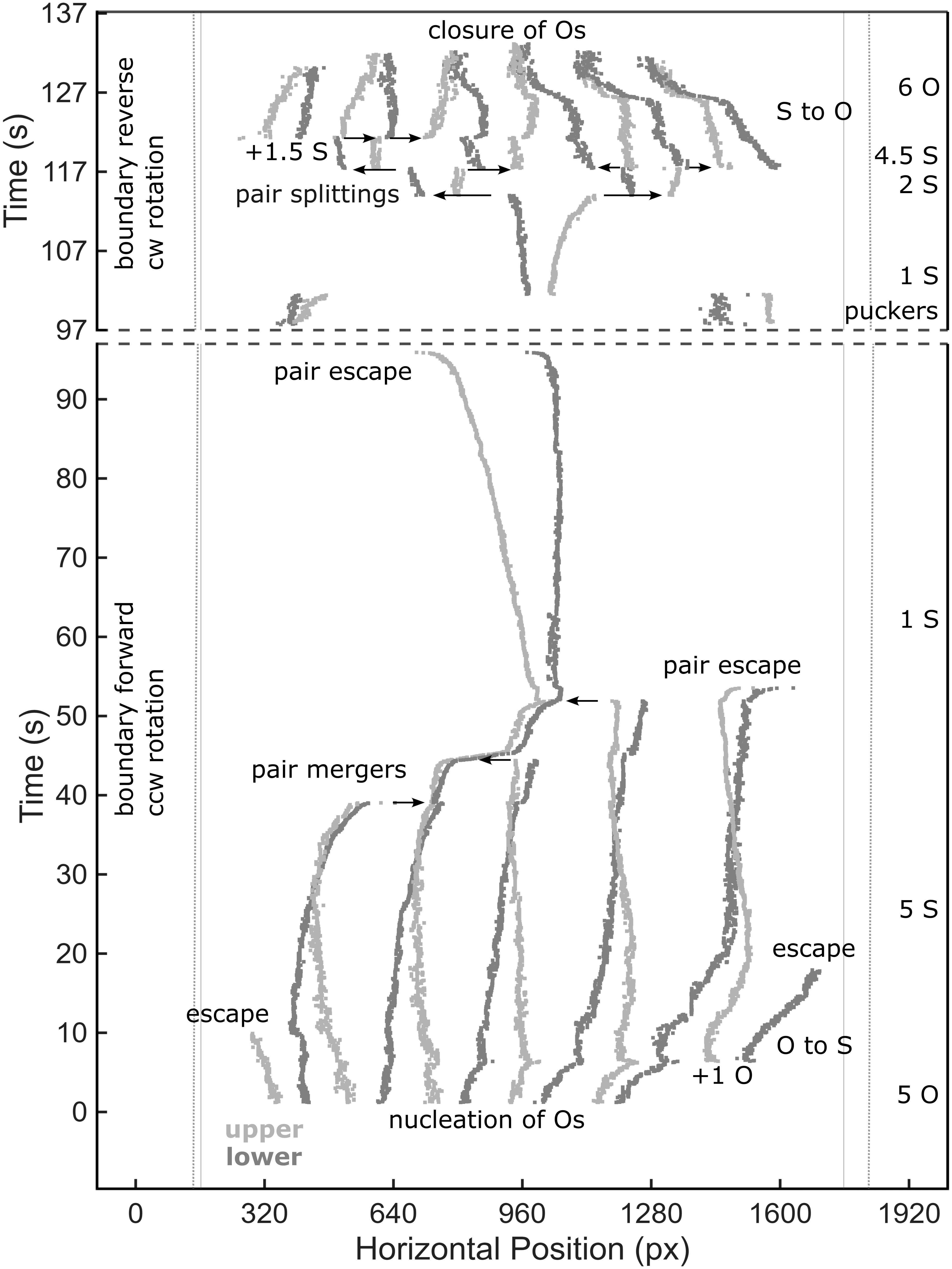}
    \caption{Spacetime plot corresponding to Video \ref{vidgreen}. Horizontal locations of upper (grey) and lower (dark grey) crumples are given in pixels, with a correspondence of approximately 135-138 px/in (5.3-5.4 px/mm) on the secant plane, with values decreasing upwards in the sheet, and unknown error due to projection of the elevated center of the sheet.  Grey lines indicate the spacetime locations of the fixed (solid) and moving (dotted, note slight slopes and jumps at the break) corners of the sheet. Some fast motions are indicated with arrows. Pair positions like the configuration at around 105 seconds correspond to a vertically oriented S-ridge. Details in text. 
    }
    \label{spacetime}
\end{figure}

\begin{video}[H]
	\centering
	\includegraphics[width=0.25\textwidth]{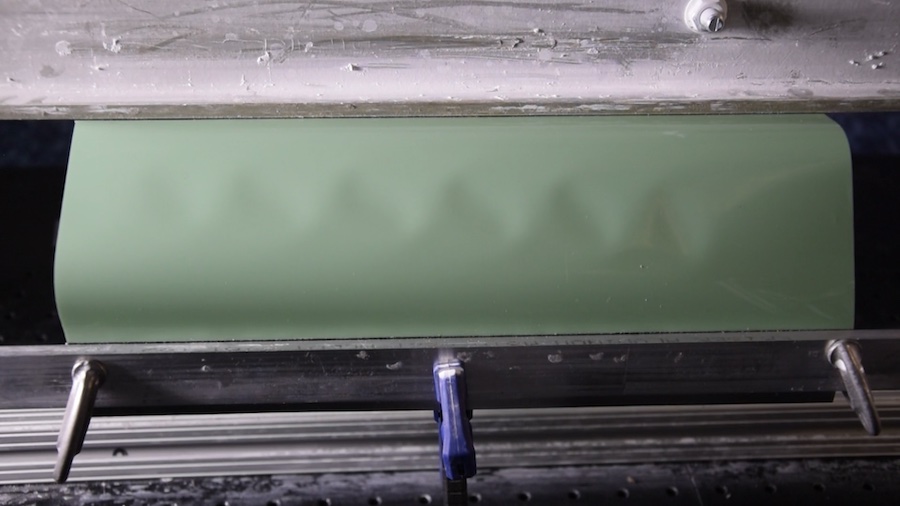}
	\setfloatlink{http://doi.org/10.5281/zenodo.18487904} 
	\caption{Patterns in 0.003 inch (0.0762 mm) polyester shim stock, 12 inches (304.8 mm) long, bent into a quarter-cylinder with 4 inches (101.6 mm) free span, clamped with lower boundary moving laterally to the right and then to the left. The boundary is moved at 0.003 mm/s and the video is at 8x speed, at 24 Hz.  Used for Figure \ref{spacetime}. (Credit: C. Redman)
}\label{vidgreen}
\end{video}

The transition from U to M and back in long sheets is as follows. 
First, a regular zipper of O-shaped oval indentations forms in rapid sequence in the U state. The crumples of this ``crystal'' then spread and subtly re-pair into a similar but more ``fluid'' array of S-shaped features, which move together as tightly bound pairs. Throughout this paper, we will refer to the O- and S-shaped pairs as ``O-valleys'' and ``S-ridges'', reflecting two different types of connection between the crumples.  Through a sequence of mergers and edge escapes, the S pattern coarsens until a single S remains, which is finally expelled through an edge. 
In this M state, small ``puckers'' can be seen, indicative of the instability of ridges (here valleys) to indentation \cite{DiDonna02}. These are potential nucleation sites for crumples, although this process has been shown to be complex, involving rapid transients in which small unstable crumples enter from inflection points in intermediate aspect ratio sheets \cite{Hutton24}. Upon reversal of the boundary motion (the video breaks and re-starts after some of this motion), a single S-ridge is nucleated, and then its two constituent crumples separate with a hint of repulsion, 
before a new O-valley is nucleated in between, leading to strong repulsion and splitting of the single S-ridge into two S-ridges.  This sequence of splitting continues, regularly or irregularly, until a full array of S-ridges returns, then subtly re-pairs into an array of O-valleys and contracts, before the valleys close, annihilating all the crumples and returning the system to the U state.

A detailed annotation of the sequence of events corresponding to Video \ref{vidgreen} is shown in Figure \ref{spacetime} below.  Several events of interest are labeled, and some states indicated on the right. Many pairwise motions of crumples, and overall rotation of the pairs due to forward and backward shear, can be seen. On top of this rotation can be seen the repulsion of the S before the first splitting event, during the return at around 110-115 seconds.  After the second splitting, an ``extra'' crumple has been nucleated near the right edge (hence ``4.5S''). The next event creates three crumples, while one S shifts rapidly rightward as indicated by the arrows; the resulting state has stray single crumples to the left and right of a 5S array (hence not ``6S''), which pattern transitions to 6O.  A similar S-pattern occurs early in the process when 5O + 1O = 6O transitions from O to S, before the two strays escape to give 5S.  Also apparent are the rapid spreading (truncated by a pair nucleation event) and contraction respectively preceding and succeeding the subtle collective transitions from O to S and S to O. Different sequences of splittings, mergers, and escapes are observed in other sheets. 

In Section \ref{results}, we further explore this and similar full loops of states, and internal loops within them, using a series of sheets of increasing aspect ratio, allowing for single or multiple cells (crumple pairs) of the pattern.

\clearpage

\section{Results}\label{results}

In this section and in Appendix \ref{additionaldata} we present force-displacement loops of various sizes for several sheets of intermediate aspect ratio, between 1.25 and 3. Each figure shows results from a single distinct sheet. All loops shown are traversed clockwise, that is, the upper branch is followed to the right and the lower to the left. Reproducibility can be gleaned from the plots. Nearly all features of the loops are highly reproducible for individual sheets.  Numbers of loops are indicated in the figure captions. The system is very sensitive to boundary conditions associated with alignment and clamping, leading to differences between sheets. 
While these experiments are an extension of the work begun in \cite{Hutton24}, we note that displacement rates here are significantly lower than the supposedly ``quasistatic'' rates used in that previous work. The sheets are also thinner and smaller, resulting in less violent release of elastic energy. These differences lead to the stable observation here of a few states only seen transiently in \cite{Hutton24}.

\begin{figure}[h]
    \centering
    \includegraphics[width=\textwidth]{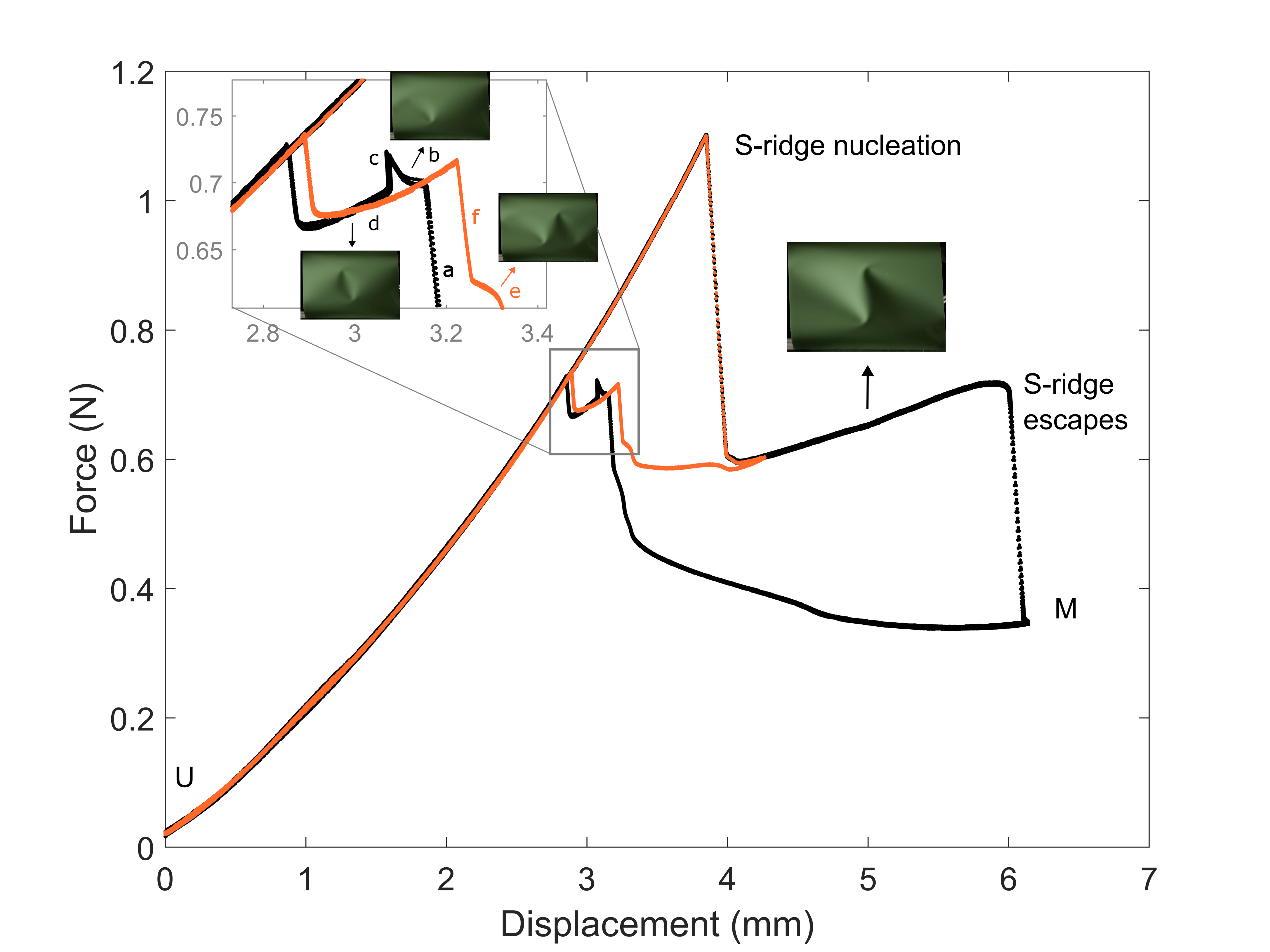}
    \caption{Force-displacement loops, traversed clockwise, for a 5 inch (127 mm) long sheet: five full loops in black and two partial loops in orange.  Some intermediate stable states are shown in the insets. Details in text. }
    \label{force_displacement_5in}
\end{figure}

Force-displacement loops for a 5 inch (127 mm) long sheet ($L/W=1.25$) are shown in Figure \ref{force_displacement_5in}, along with insets showing several intermediate stable states of interest.  This sheet is long enough to admit about one unit cell of the pattern, consisting of two crumples in either an O or S configuration. The five black curves, which sit very closely on top of each other, correspond to a full loop between U and M states.  The force increases monotonically with displacement until a large drop corresponding to the first bifurcation from the U state, resulting in a single S-ridge. The force increases monotonically again until the final bifurcation expelling the S-ridge from the edge of the sheet, resulting in complete snap-through to the M state. Reversing the boundary shear and returning along the lower branch, there is a jump in force (a) to a short-lived single-crumple state (b), soon followed by a drop in force (c) corresponding to the edge-entrance of another crumple, which binds to the first to form an O-valley (d). 
The O-valley shrinks until it collapses with a final jump in force, bringing the sheet back onto the original U state branch. The two orange curves, also in very close agreement, correspond to a partial loop between U and a point shortly after the first bifurcation to an S-ridge. Upon return, the S-pair slowly rotates clockwise, until a new crumple enters from the lower right (e).  During the jump (f), the upper crumple in the S-pair re-pairs with the newly entered crumple to form an O-pair, and then the lower crumple in the S-pair is expelled from the left edge, resulting in a single O-valley, whose closure returns the system to the original branch. This last intermediate process on the inner loop (orange) does not superimpose on the corresponding part of the full loop (black), as the O-valleys in these cases are not identically shaped or situated. 

\begin{figure}[h]
    \centering
    \includegraphics[width=\textwidth]{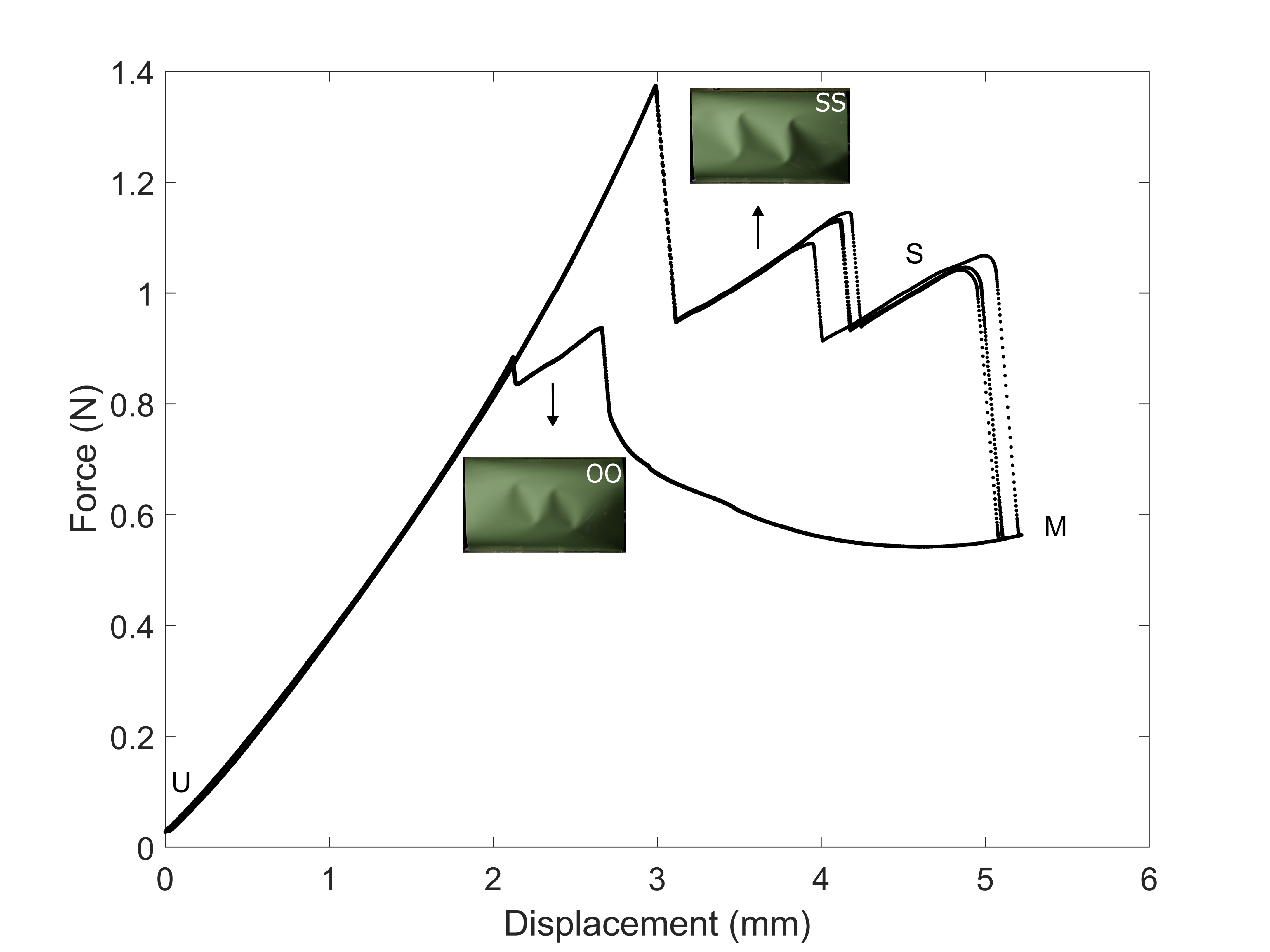}
    \caption{Four force-displacement loops, traversed clockwise, for a 6 inch (152.4 mm)  long sheet: increasing shear in red and decreasing shear in blue.  Insets show stable states with two pairs of crumples in S-ridge and O-valley arrangements. Details in text. }
    \label{force_displacement_6in}
\end{figure}

Four full force-displacement loops for a 6 inch (152.4 mm) long sheet ($L/W=1.5$) are shown in Figure \ref{force_displacement_6in}, along with two insets. 
This small increase in sheet length provides enough room for an additional cell of the pattern. The first bifurcation and force drop results in a pair of S-ridges.  These rotate until the next drop when one ridge is expelled from the edge. The remaining ridge rotates until it is expelled in the final bifurcation. On the return branch, a pair of O-valleys forms. These close essentially simultaneously. Most features of the loops are highly reproducible, but the displacement at which S-ridges are expelled can vary. 

\begin{figure}[h]
    \centering
    \includegraphics[width=\textwidth]{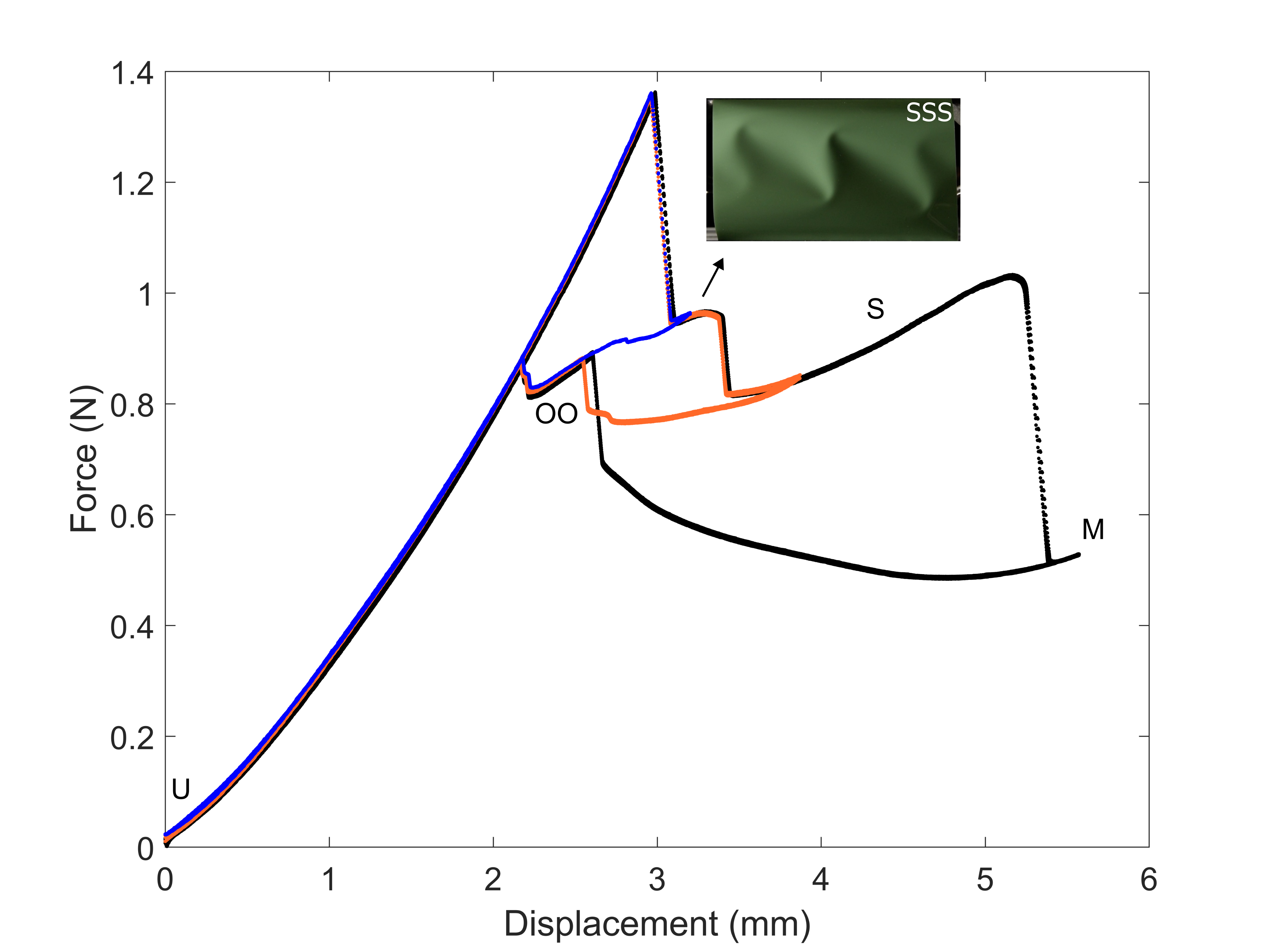}
    \caption{Force-displacement loops, traversed clockwise, for a different 6 inch (152.4 mm) long sheet: four full loops in black and two partial loops each in orange and blue.  Here the first stable state observed has three S-ridge pairs, as shown in the inset. Details in text. }
    \label{force_displacement_6in_alt}    
\end{figure}

Force-displacement loops for a different sheet of the same size are shown in Figure \ref{force_displacement_6in_alt}, along with an inset. Here a triple-S pattern forms instead of a double-S pattern, and exists over a shorter interval. This difference between the two sheets is an example of how sensitive this type of system is to boundary conditions, and how inertial effects from released energy can cause the system to jump past stable states.   
Here the four black curves are full loops, while the two each of orange and blue curves are partial loops between U and the intermediate S and triple-S states, respectively. The orange S branch shows a bit of hysteresis, with return deviating from forward. The creation of OO occurs through a re-pairing after nucleation of a crumple on the lower right (small shoulder) and then another on the upper left (large jump). This process can be compared with the behavior corresponding to e-f in Figure \ref{force_displacement_5in}, described previously. On the blue branch, the SSS to OO transition occurs through expulsion of crumples on the upper right and lower left. 

\clearpage

\begin{figure}[h]
    \centering
    \includegraphics[width=\textwidth]{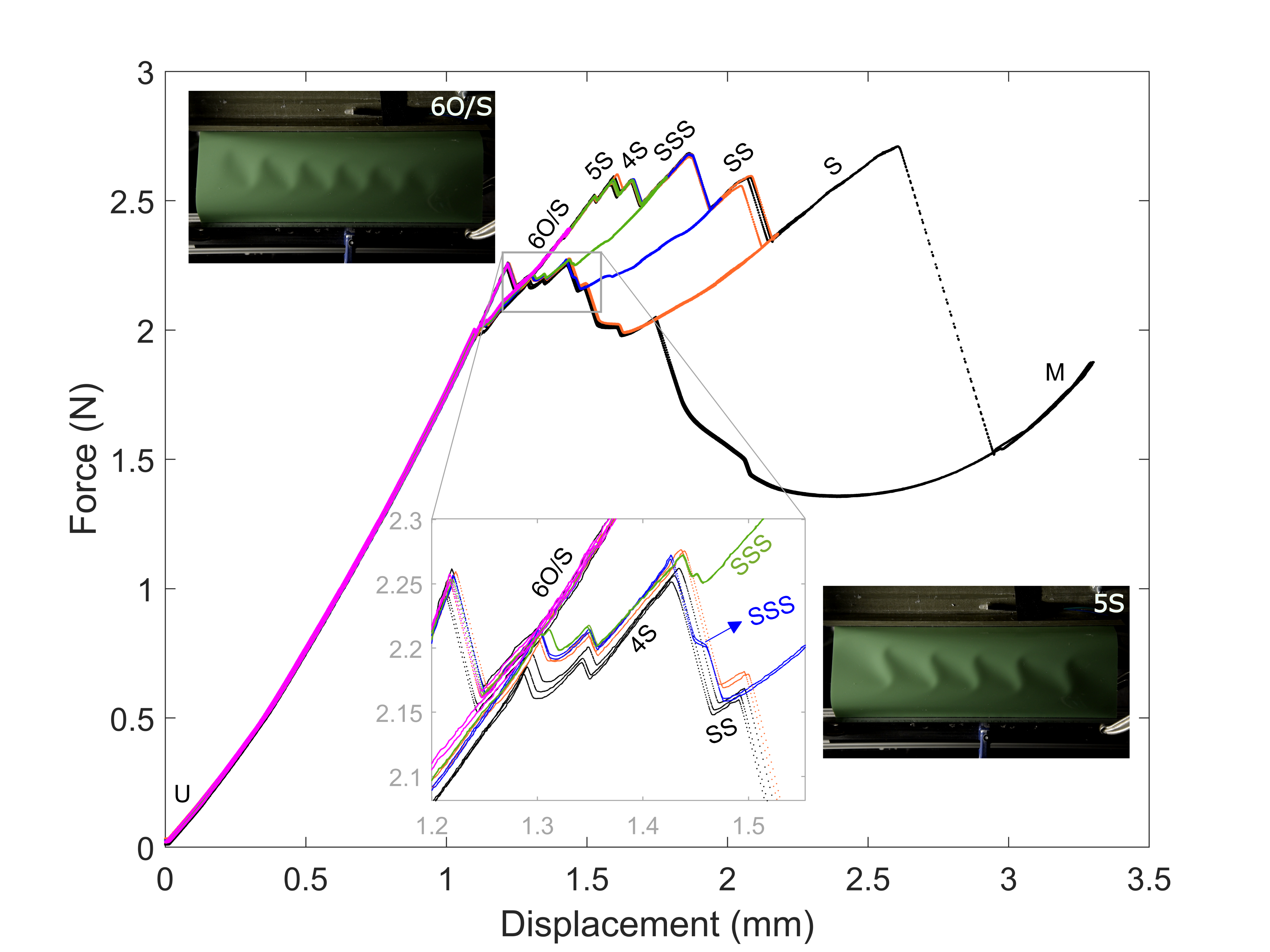}
    \caption{Force-displacement loops, traversed clockwise, for an 11 inch (279.4 mm)  long sheet: three full loops in black and two partial loops each in orange, blue, green, and pink. Details in text. 
    }
    \label{utom11}
\end{figure}

Force-displacement loops for an 11 inch (279.4 mm)  sheet ($L/W=2.75$) are shown in Figure \ref{utom11}. This longer sheet has room for more unit cells. Here the three black curves are full loops, while the two each of orange, blue, green, and pink curves are partial loops between U and several other states. The curves are complex but reproducible, with signatures of a many small events. The transition from O to S pairing is too subtle to distinguish. Higher-number patterns are harder to access and interpret, likely because of released kinetic energy as well as interactions with the free edges, such as the induction of crumples. These effects also interfere with the ideal doubling sequence. 
However, it is clear that while going both forward and backward, the system visits identifiable branches containing one, two, three, four, and possibly more pairs of crumples that exist in overlapping regions of the boundary displacement. The observed branches have positive slopes. 
The least reproducible aspect of the curves is the transition from SS to S, which on this sheet proceeds through a merger rather than an escape.  For this sheet, the coarsening sequence from 5S was: escape of rightmost S, escape of leftmost S, merging on the left, merging, escape.

Additional full and partial loop force-displacement data for various sheet lengths is included in Appendix \ref{additionaldata}.  While they do not directly correspond, the responses shown in Figure \ref{utom11} and throughout the Appendix can be compared with the annotations of Video \ref{vidgreen} in Figure \ref{spacetime}. 

\begin{figure}[h]
    \centering
    \includegraphics[width=\textwidth]{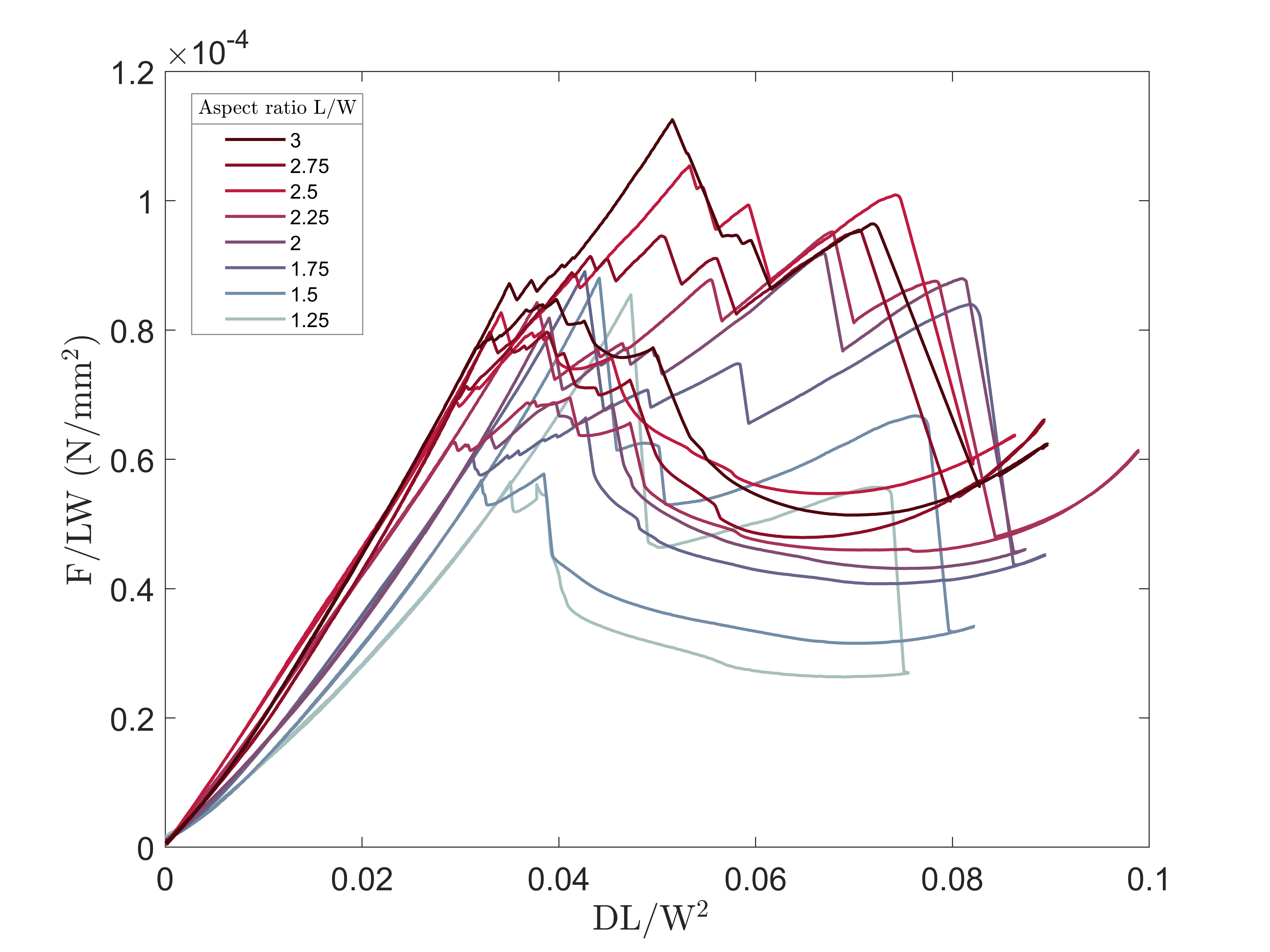}
    \caption{Full force-displacement loops for sheets of lengths 5 to 12 inches (127 to 304.8 mm, aspect ratios 1.25 to 3), with displacement $D$ divided by the width $W$ and multiplied by aspect ratio $L/W$, and force $F$ divided by sheet area $LW$.  Displacement zeroing is approximate.}
    \label{overlay}
\end{figure}

Figure \ref{overlay} overlays full force-displacement loops for sheets. For clarity of visualization, we select the penultimate loop from each set of loops, and use only one set for each aspect ratio.  
 The displacement has been divided by the width $W$ and then multiplied by aspect ratio $L/W$, because the normalized displacements at which bifurcations occur scale inversely with aspect ratio at large aspect ratios \cite{YuHanna19, Hutton24}.  This scaling reflects geometric constraints on the range of motion of the sheet edges for inextensible deformations of the sheet. 
The force has been divided by the total area of the sheet $LW$, as we expect the global bending resistance to scale with the size of the sheet, for a fixed thickness. We report this quantity as a dimensionful stress. 
Recall that neither $W$ nor thickness are varied in these experiments. 
  These simple scalings bring some aspects of the plots into rough correspondence, particularly for the longer sheets. 
The point of this rescaling is to present the data comparably in a single plot. As the sequence of deformations differs between different aspect ratio sheets, no collapse is expected except the initial slope, the displacement and force at the first bifurcation, and possibly the displacement at the last bifurcation, and all only for large aspect ratios.

\begin{figure}[h]
    \centering
    \includegraphics[width=\textwidth]{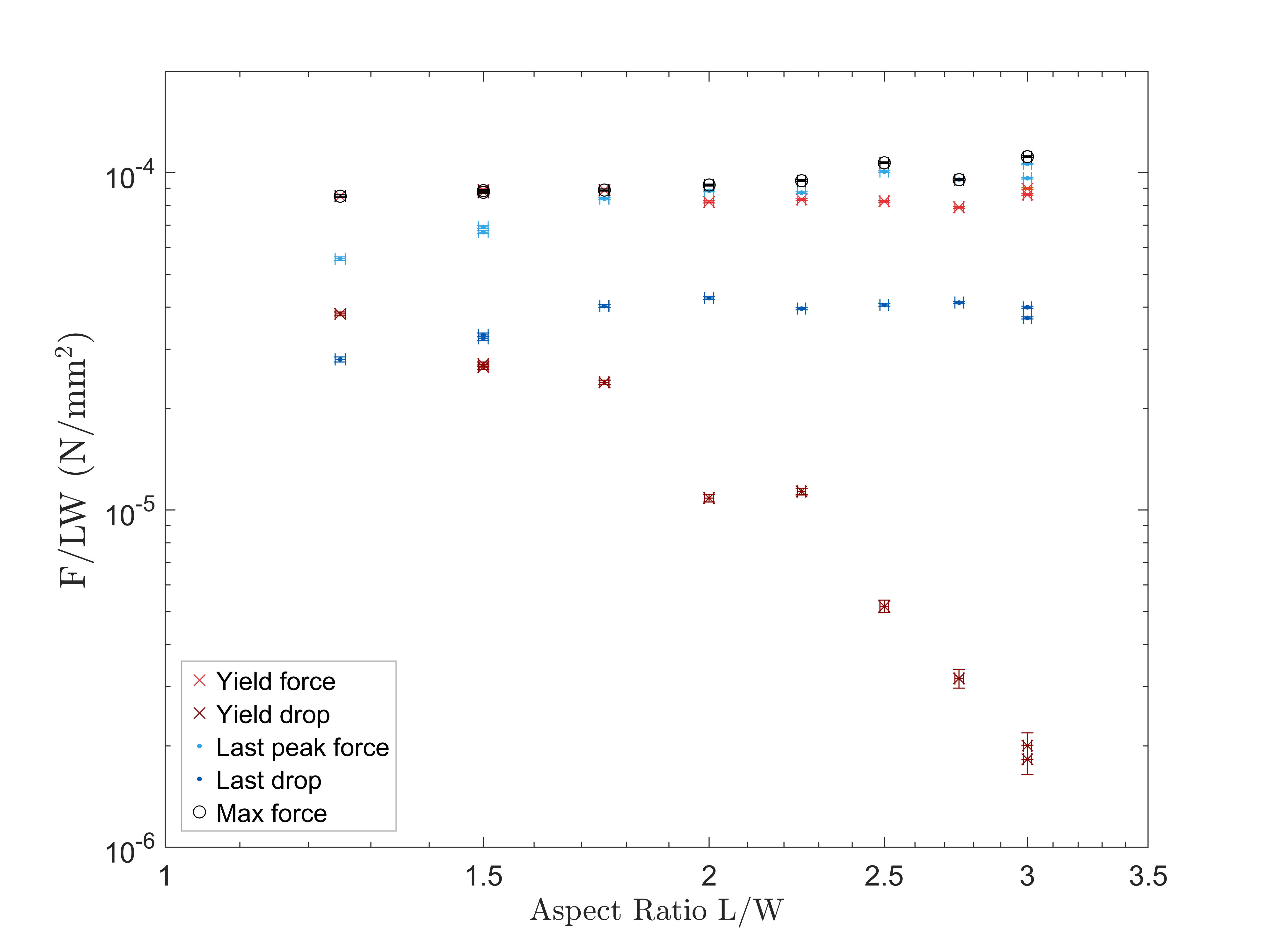}
    \caption{Forces (divided by sheet area) corresponding to the loops used in Figure \ref{overlay} and two additional loops: the first (yield) force peak and corresponding drop in force, the last force peak and corresponding drop, and the maximum force, which is associated with different features among the curves, and at times coincides with other forces shown. Error bars denote uncertainty of individual measurements from noise (for force drops), and noise and drift (for forces), and uncertainty in geometric quantities used for normalization. These data use the penultimate loop from each set; variation between loops in a set is consistent with the error estimates.} 
    \label{forces}
\end{figure}

Basic information about forces is extracted from these loops and two additional loops, and plotted on a log scale in Figure \ref{forces}. 
This shows five relevant forces, divided by sheet area: the first (yield) peak and drop arising from the first bifurcation giving rise to the first O-valley(s), the last peak and drop from the escape of the last S-ridge to achieve final snap-through, and the maximum force, which for shorter sheets corresponds to an early peak and for longer sheets corresponds to later peaks, although these are not that different in value. 

Several of the forces, in particular the last drop in force, approach flatness on this plot for longer sheets, meaning that they simply increase linearly with length (equivalent to area here, as we have not varied the width).  While the yield force over length (area) is roughly flat, the yield drop decreases rapidly with sheet length, reflecting a more gentle response at the first bifurcation, with higher loads after restabilization, for longer sheets in this intermediate range of aspect ratios.

\section{Discussion}\label{discussion}

Our results provide experimental evidence for snaking in buckled plates. 
A system exhibiting snaking has two intertwined branches of solutions corresponding to odd and even numbers of unit cells of a pattern that can coexist with a uniform state of similar energy. Each branch has stable segments connected by unstable retrograde segments between two folds, and additional unstable asymmetric states can join these branches to form a ladder-like structure \cite{Dawes10, Knobloch15}. 
The utility of this concept in understanding pattern formation in buckled structures has been demonstrated in recent works employing finite element or PDE representations of shallow shells \cite{KreilosSchneider17, GrohPirrera19, Hunt20, Groh21, GrohHunt21}, and the present work suggests possible starting points for path following employing such methods. 
These works compared numerical solutions with experimental observations of sequences of events reported decades ago by Yamaki 
and/or E{\ss}linger and Geier, 
with the benefit of both advances in numerical continuation methods, and the theoretical perspective provided by the snaking framework as developed over the last quarter of a century, well after the original experiments. 
Our plots also show similarities to those obtained by Yamaki \cite[figures 3.47, 3.52b-e]{Yamaki} and E{\ss}linger and Geier \cite[figure 3.21]{EsslingerGeierBook}. 
Figures such as \ref{force_displacement_6in_alt} and \ref{utom11} show multiple overlapping branches, each containing a different number of crumple pairs.  
One can also compare with the experimental results of Ravulapalli and co-workers \cite[figures 9-14]{Ravulapalli24}, who did not trace the response along internal return branches. This work also compared with finite element results, and invoked the snaking framework. 

Within this framework as applied by these previous authors, the unit cell of the patterned state is what we here call an O-valley, and the two branches that ``snake'' around each other comprise odd and even numbers of O-valleys.   Formation of an O-valley connects the two branches along a rung of a ``ladder'' \cite{Groh21}. However, these other experiments were performed on cylindrical structures with quite different boundary conditions than the present work. Instead of a complete, closed cylinder under compression and/or twist and/or internal pressure, we applied a shear deformation to a quarter-cylinder with open ends. 
Many of our resulting observations involve pairs in the form of S-ridges, but arrays of these are closely related to arrays of O-valleys, and the splitting of such a ridge into two ridges involves the nucleation of an O-valley. 
It is also important to note that much experimental work on metallic or polymeric ``cylindrical shells'' was actually performed using bent sheets with a lap joint \cite{Yamaki, Harris61} or butt strap joint \cite{EsslingerGeierBook}, while some used structures with a natural cylindrical state created by electrodeposition \cite{Almroth64}, spin-casting \cite{Tennyson69}, or 3D-printing \cite{Ravulapalli24}.  This distinction between bent plates and proper shells is often overlooked, and its consequences remain unclear. Much of the experimental literature uses the term ``shell'' indiscriminantly to mean both. Recent works applying snaking concepts to earlier experiments have employed shallow shell equations or shell finite elements \cite{KreilosSchneider17, GrohPirrera19, Groh21, GrohHunt21, Ravulapalli24}. 
Of these, only the case of Ravulapalli and co-workers is a comparison of experimental and simulated shells. 

Following unstable paths to determine the connectivity of states would provide stronger support for the presence of snaking in this system.  This could be achieved with a computational campaign like those described above, or possibly with an experimental probing and control method like that developed in \cite{Neville20}. 
The latter would require adaptation to allow for probes that can freely move in two directions, either in a plane, or preferably circumferentially and axially with respect to the unsheared state. 

Branches with five or more pairs, and O-pair regimes that exist in a small range of lower shears, are a bit more ambiguous to observe due to edge effects adding/subtracting individual crumple half-units, and are harder to access with internal loops in these sheets. There is additional complexity here because of the subtle transitions between O and S pairs, although we note interesting transitions were shown in very short cylinders in \cite[figure 3.44b]{EsslingerGeierBook}. 
Furthermore, some of the subtle interactions between crumples can be attributed to the formation and breaking of ``bonds'' that form the subject of 
 another paper \cite{Gimenopairs}. 
A portion of the return branch in longer sheets, corresponding to the breaking of an S-ridge just before splitting, is measured more carefully there. 
 The trends seen in these intermediate aspect ratio sheets were also informally observed at higher aspect ratios that allow for more unit cells.  However, with these dimensions come greater difficulties in accurately setting boundary conditions, and smaller displacement ranges and jumps in force. 
  
Across all the experiments, exceptions to reproducibility are relatively rare, and usually involve the timing of unstable events such as splitting, merging, and escape.  Factors negatively influencing reproducibility include larger energy releases in shorter sheets and an increased difficulty of alignment in longer sheets, which also admit more possible configurations and transition sequences.  

Our system has allowed us to observe several mechanisms for coarsening and refinement of the pattern. Of these the most interesting is the splitting of the S-ridge, which involves a bond breaking and weak repulsion between individual crumples, 
 then nucleation of an O-valley (related to the instability of ridges to indentation \cite{DiDonna02}) and concomitant strong repulsion between the resulting crumple S-ridge pairs. It is unknown whether such behaviors are system-specific, or whether something analogous might be observed more generally across model pattern forming systems. 
Merging of S-ridge pairs was also observed. 
Other mechanisms, including escape, were enabled by the open boundaries of our system. Crumple induction events near boundaries might be compared with similar events in the interior \cite{Marthelot17, WittenMovsheva25}. 

The classic O-valley pair  is often seen as the unit of static patterns on cylinders.  While we observe these here in our open-boundary system, the highly mobile but tightly bound S-ridge pair appears to be the dominant motif enabling transitions between patterns and mediating the global snap-through of the structure. 
Our different perspective on this class of problems comes from a line of thinking in which a single crumple is the relevant unit of both static and dynamic deformations, and the O and S are two of several possible types of pairs that these anisotropic units can form.

Most forces in these sheets increase simply with sheet length or, equivalently here, area. For example, the final drops in force suggest that the energy of an isolated S-ridge in a long sheet is associated with a nearly-constant cross section away from the ridge.
However, the initial yield drops show a rapid decline with length. Yield corresponds to the nucleation of O-valleys, which in longer sheets are shallower and play a more incremental role with respect to the full snap-through, which still requires a force that scales with length.

\section{Conclusions}\label{conclusions}

A sequence of bifurcations and patterns in an elastic sheet bring part of a cylinder to a snapped-through configuration and back again, with an S-shaped pair motif playing an important role.  Partial loops reveal behavior consistent with snaking.  Mechanisms enabling these transitions, including splitting, merging, and escape of S-pairs, were discussed, along with some aspects of the forces required.

\section*{acknowledgments}
We thank R. Groh for many helpful discussions before and during the writing of the paper, T. Yu for early discussions and ideas, B. Nagy for machining cantilevers, C. Redman for help with videos, and A. Robinson and J. Rudfelt for help with experiments.  
JAH acknowledges support from the U.S. National Science Foundation under grant CMMI-2210797. 
EH is grateful for the support of DICYT (USACH) under grant 042431HH. 

\section*{additional files}
\noindent Videos: \url{http://doi.org/10.5281/zenodo.18487904}  \\
\noindent Data: \url{https://doi.org/10.5281/zenodo.18487934} 

\section*{author contributions}
EH and JAH conceived and guided the study. DG and EH designed the apparatus. 
DG performed the experiments. 
GF and RSH contributed to the apparatus and performed early exploratory experiments. 
DG, BKM, EH, and JAH analyzed data and made figures.  
JAH wrote the first draft, and DG and EH contributed to the final draft.  

\bibliographystyle{unsrt}

\appendix

\newpage

\section{Additional data}\label{additionaldata}

\begin{figure}[H]
    \centering
    \includegraphics[width=0.725\textwidth]{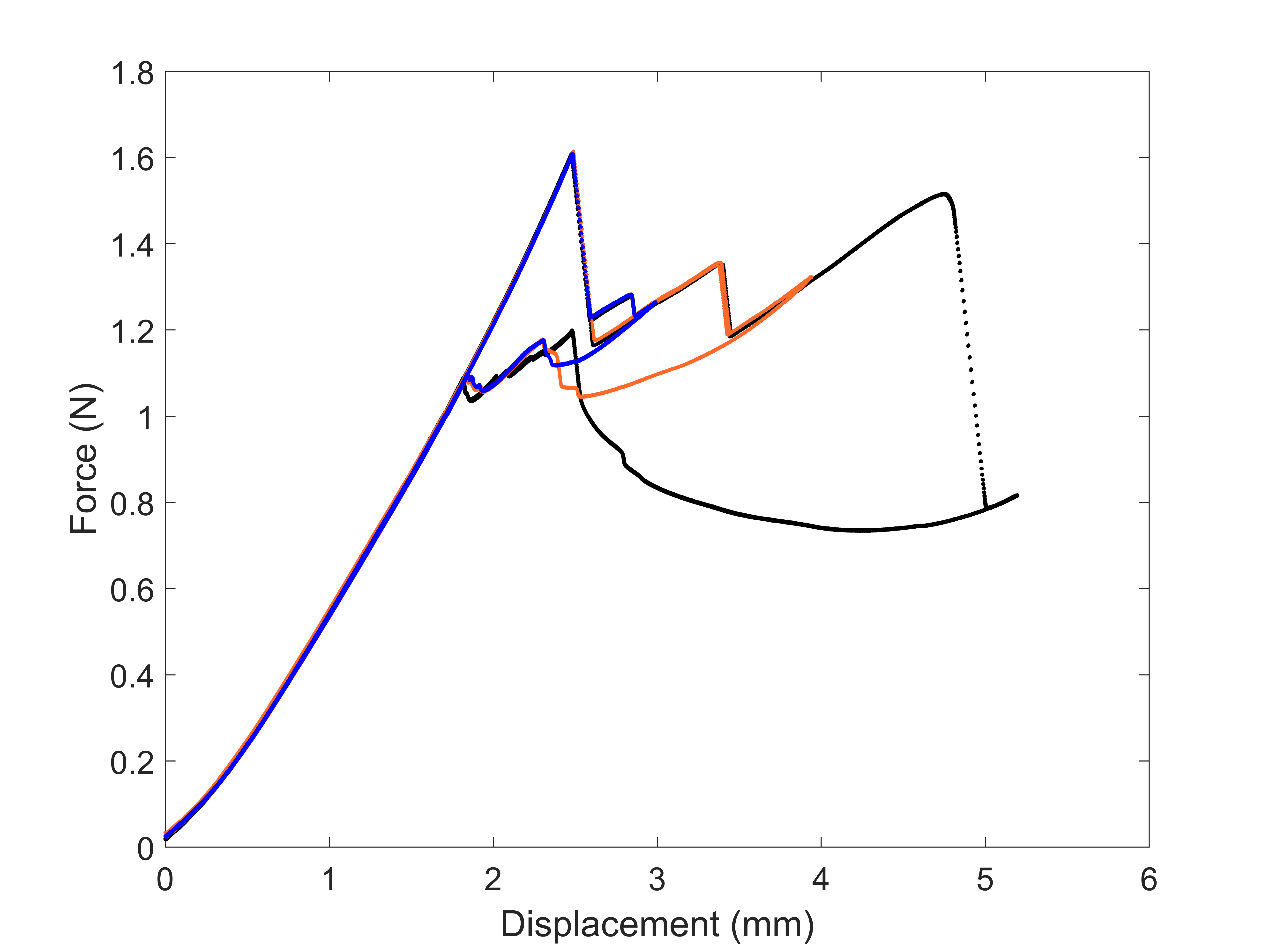}
    \caption{
Force-displacement loops, traversed clockwise, for a 7 inch (177.8 mm) long sheet ($L/W=1.75$): three full loops in black and two partial loops each in orange and blue. 
    }
    \label{utom7}
\end{figure}

\begin{figure}[H]
    \centering
    \includegraphics[width=0.725\textwidth]{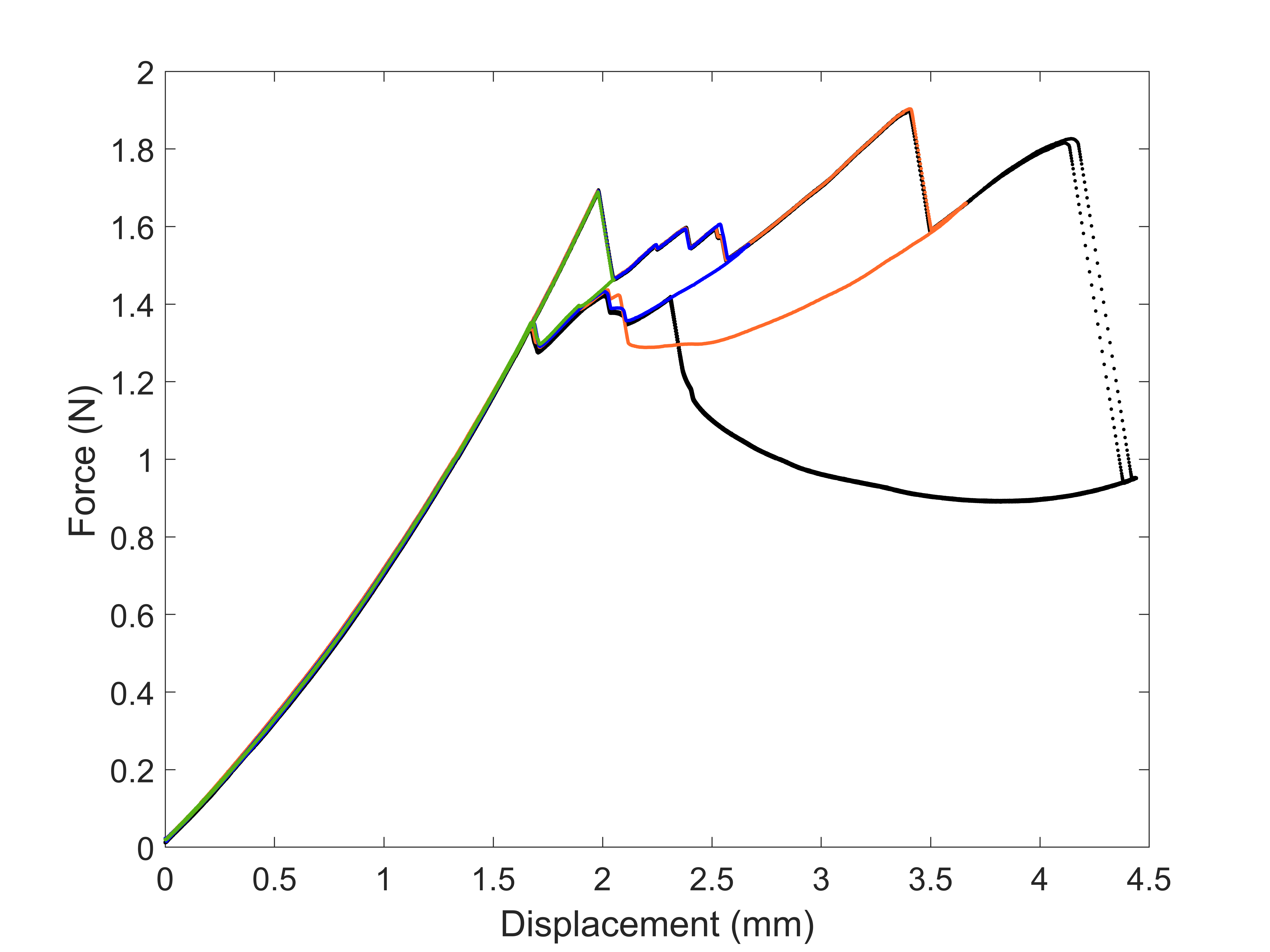}
    \caption{
Force-displacement loops, traversed clockwise, for an 8 inch (203.2 mm) long sheet ($L/W=2$): two full loops in black and one partial loop each in orange, blue, and green. 
    }
    \label{utom8}
\end{figure}

\begin{figure}[H]
    \centering
    \includegraphics[width=0.725\textwidth]{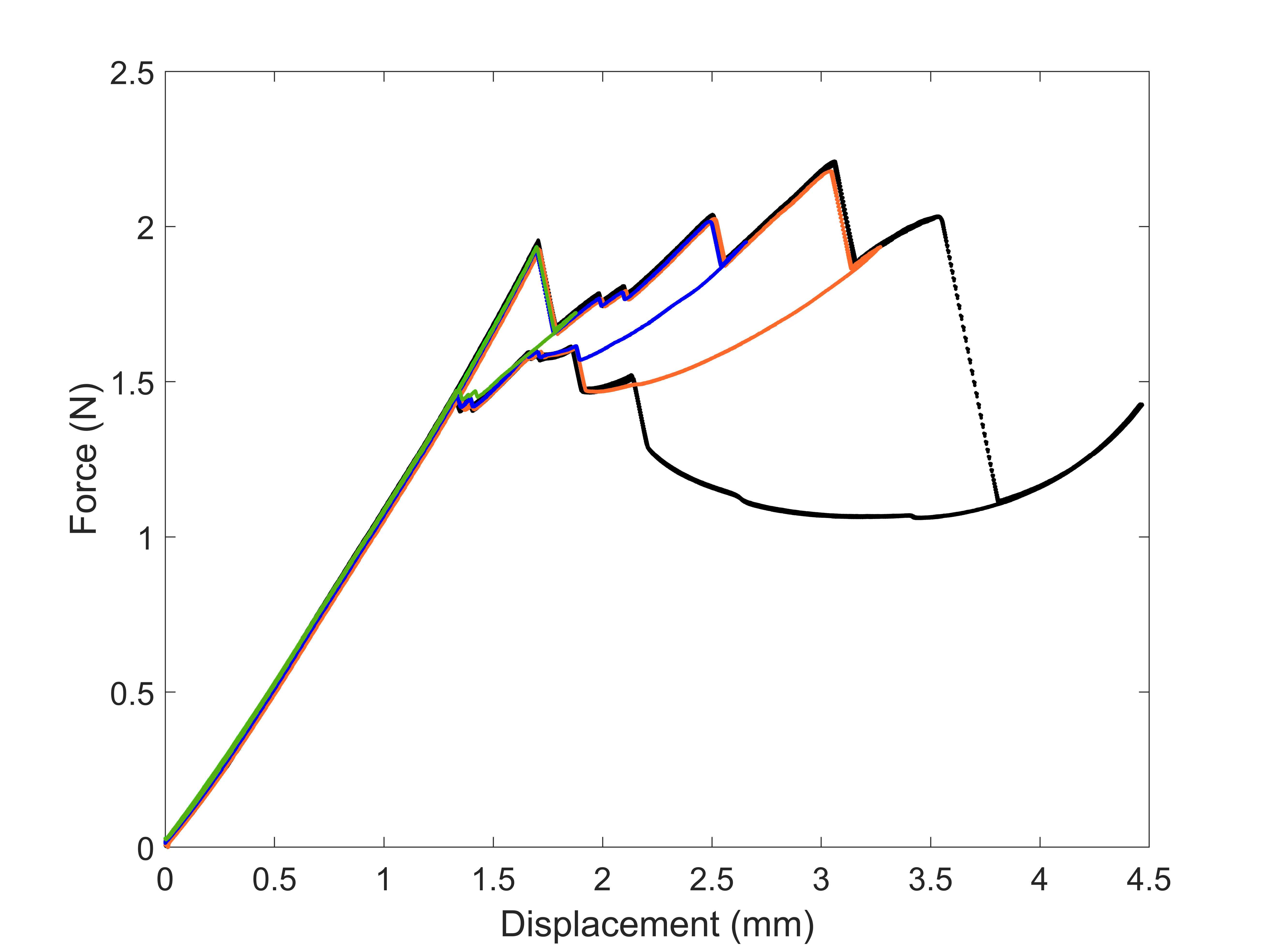}
    \caption{
Force-displacement loops, traversed clockwise, for a 9 inch (228.6 mm) long sheet ($L/W=2.25$): four full loops in black and two partial loops each in orange, blue, and green. 
    }
    \label{utom9}
\end{figure}

\begin{figure}[H]
    \centering
    \includegraphics[width=0.725\textwidth]{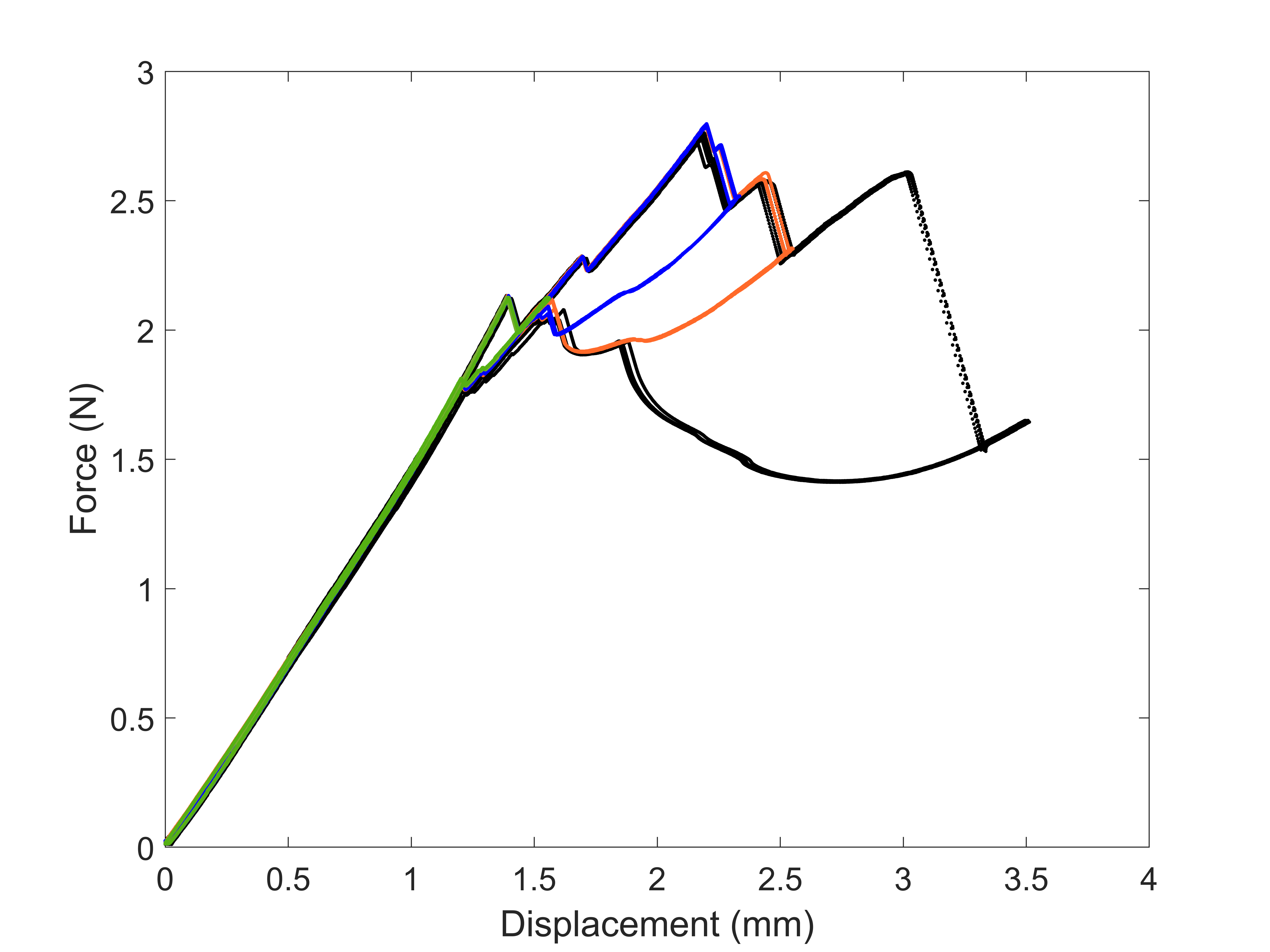}
    \caption{
Force-displacement loops, traversed clockwise, for a 10 inch (254 mm) long sheet ($L/W=2.5$): five full loops in black and three partial loops each in orange, blue, and green. 
    }
    \label{utom10}
\end{figure}

\begin{figure}[H]
    \centering
    \includegraphics[width=0.725\textwidth]{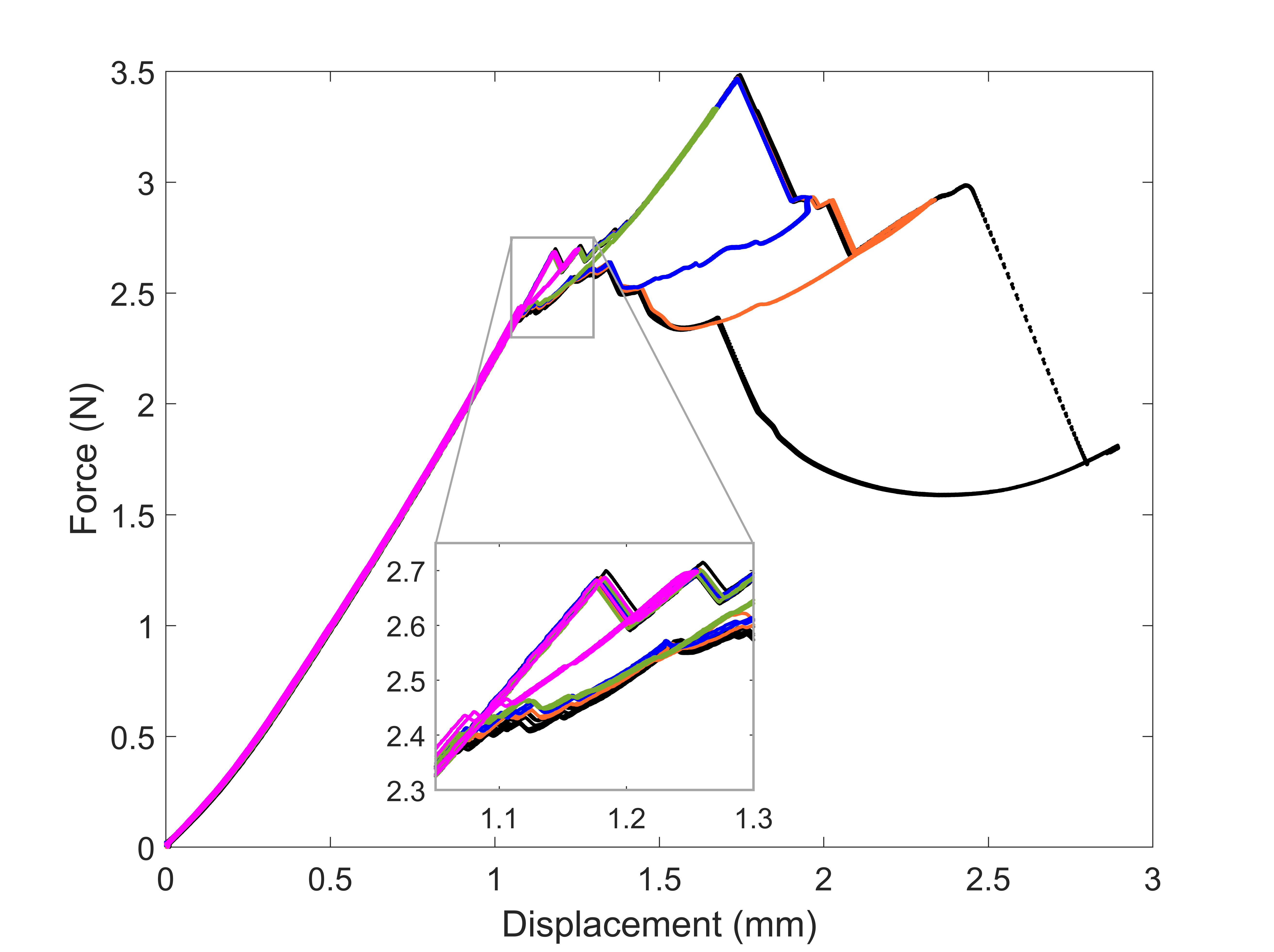}
    \caption{ 
Force-displacement loops, traversed clockwise, for a 12 inch (304.8 mm) long sheet ($L/W=3$): five full loops in black and two partial loops each in orange, blue, green, and pink.  
    }
    \label{force_displacement_12in_inner}
\end{figure}

\begin{figure}[H]
    \centering
    \includegraphics[width=0.725\textwidth]{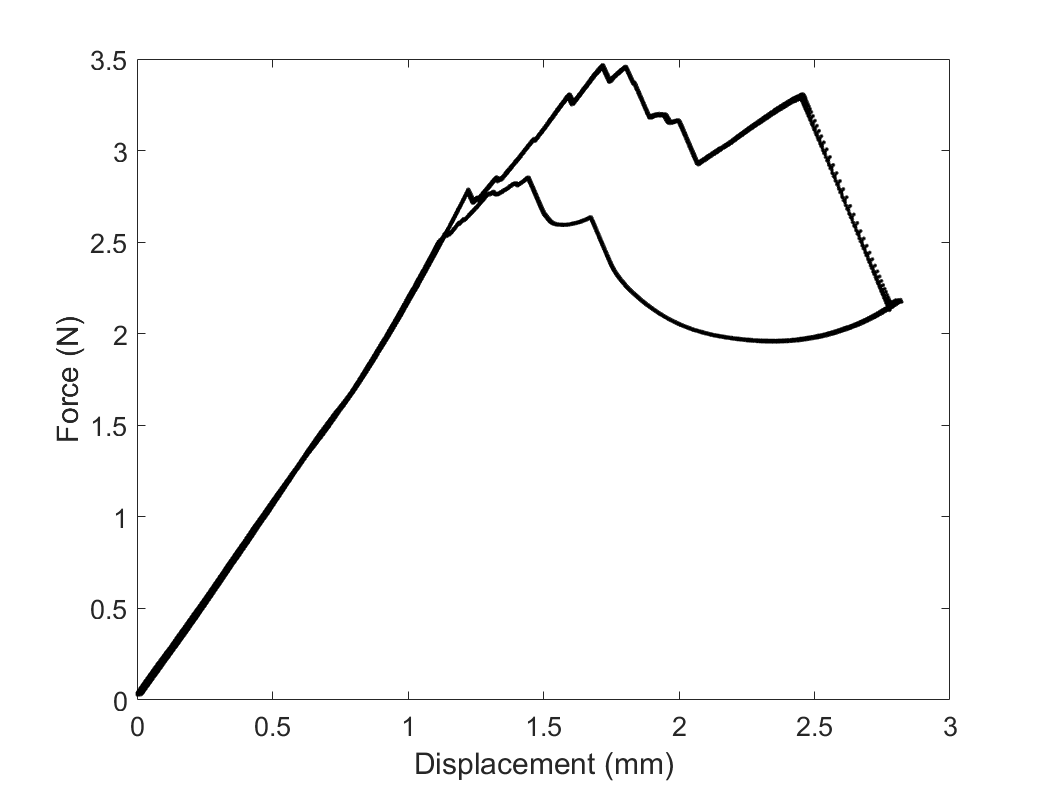}
    \caption{Six force-displacement loops, traversed clockwise, for a different 12 inch (304.8 mm) long sheet than that of Figure \ref{force_displacement_12in_inner}. Note different behavior near the highest peak. For this sheet, the coarsening sequence from 5S was: escape of leftmost S, escape of rightmost S, merging on the left, merging, escape. On the return branch, the single S jumped directly to 4S. Note differences with the sequence for the different 12 inch sheet in Figure \ref{spacetime} (which was also deformed at a slower speed) and similarities with most of the coarsening sequence for the 11 inch sheet in Figure \ref{utom11}. 
    }
    \label{force_displacement_12in}
\end{figure}

%
%
%
%
%
%
%
%
\end{document}